# Wireless Data Link at 1Gbps using 256 QAM

David Noel
Department of Electrical and Computer Engineering, University of Florida

**Abstract**

*This report describes the design and proposal of a wireless link capable of broadcasting at 1 Gbps. For this application, isotropic antennas, 256 QAM modulation, and BER level less than 1e-5, without using error correction coding, were implemented. A frequency of 5GHz was employed to achieve such high data rates. For unlicensed operations in this frequency range, the FCC allocates a 5.15 - 5.35 GHz frequency range with maximum acceptable power levels no greater than 250mW(~24dBm)[2]. Due to its inexpensiveness and simplicity, the transceiver architecture and all its subsystems used the homodyne system. The complete system architecture is described with some of their most significant performance characteristics, including modulation, fundamental and 3rd harmonics, power spectra, and constellation diagrams. To conclude, a Bill of Materials (BOM), costs, and associated specifications were included.*

**Keywords:** Gbps wireless link, BER, 256 QAM, homodyne architecture, harmonics.

## 1. Introduction

The need for faster and more efficient wireless data transmission rates has been exponentially growing due to the need of consumers to communicate large data streams, such as 1080p video streaming, faster, securely, and over far distances. These tasks hastened the current utilization of the 802.11n wireless standard(300Mbps), but this data rate is now becoming a bottle-neck and faster, more secure data rates are requisite, which introduces the 802.11ac standard (Gbps data rates).

The approach to designing this system uses 256 QAM modulation, which provides very high data-rates using bandwidth-limited channels and integrates very well with current technologies[3]. From the issues presented, one can see the need for "breaking" into the Gbps range and engender the widespread use and adoption of 802.11ac.

## 2. Transceiver Design

Table 1 outlines the specifications and commonalities used in the design of the Transmitter and Receiver. Consideration was not given to multiple access and duplexing, just TX-to-RX free space transmission and polarization is matched.

| Transmitter | Receiver | Shared |
|---|---|---|
| LO Frequency: 5.1 GHz | Input Frequency: 5 GHz | Antenna: Isotropic |
| Power Out: 23.31dBm | Noise Figure(dB): 6.24 | Modulation: 256 QAM |
|  | Sensitivity(dBm): -54.37 | Bandwidth: 250 MHz |
|  |  | Rate: 1 Gbps |

**Table 1.** Specifications of the Transmitter and Receiver.

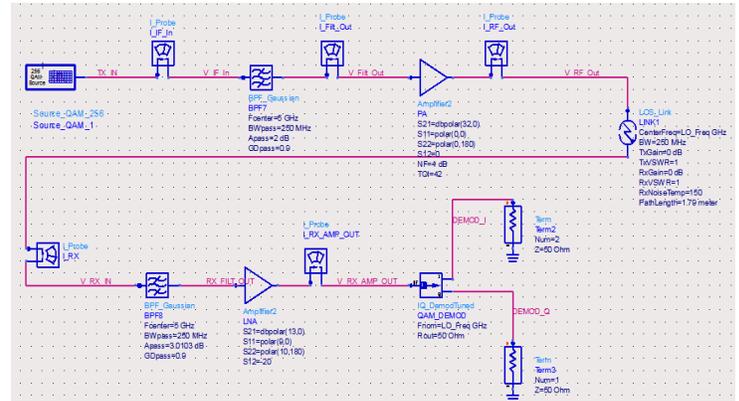

**Figure 1.** Transceiver System Schematic.

### 2.1 Transmitter

Homodyne architecture (direct up-conversion) was employed for the transmitter design. The DAC was emulated using 8 random bit-sequence generators (4 per channel) and modulated using a Local Oscillator and an ADL5375 quadrature modulator by mixing baseband with 5 GHz and the in-phase (I) and quadrature phase (Q) channels with a 90-degree phase shift. After modulation, the signal was filtered using a Gaussian bandpass filter due to its minimization of rise-fall time and group delay. After being filtered, the signal entered a power amplifier (PA) and was amplified to a 23 dBm signal for transmission.

This system provided a high level of integration since only one LO was used. The produced signal, however, was disadvantaged by the large LO pulling by the PA, which is inherent in Homodyne systems.

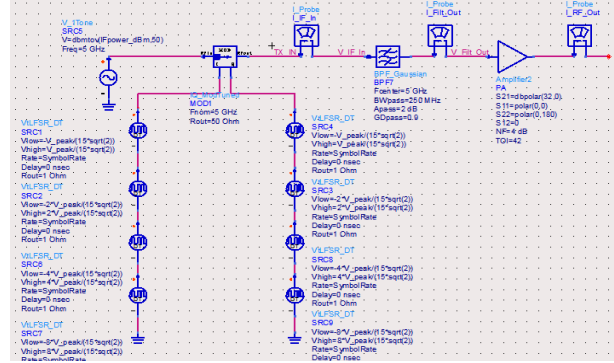

**Figure 2.** Transmitter Schematic.

### 2.2 Receiver

Parallel to the Transmitter, a Direct Conversion Zero-IF architecture was chosen for the Receiver. The input signal was first filtered using a Gaussian Bandpass filter, and then the signal was passed through a low-noise power amplifier (LNA). After amplification, the signal was demodulated to a baseband using an ADL5380 quadrature demodulator. After demodulation, the signal was divided into its respective I and Q channels and then output to its respective ADC system. This system offers a very high level of integration and the absence of the image frequency problem. However, inherently present and limiting are DC coupling and injection locking.

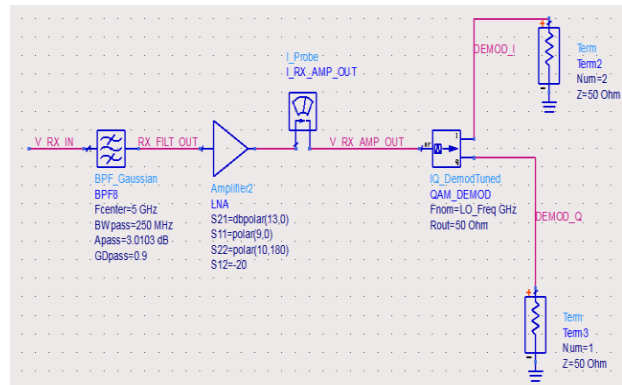

**Figure 3.** Receiver Schematic.

### 3. Modulation Scheme

The modulation chosen for this design was a 256 QAM scheme since it has high spectral efficiency, provides very high data rates using bandwidth-limited channels, and integrates very well with current technologies. With this scheme, a symbol represents 8 bits of data ($N = \frac{8\,bits}{sym}$), which results in a Bandwidth, $BW = \frac{R}{N} = \frac{1Gbps}{8bits} = 125 MHz$. This result, in turn, leads to a null-to-null or channel bandwidth of $B_{null} = 2 * BW = 2 * 125 MHz = 250\ MHz$. However, due to its transmission of more bits per given bandwidth or $\frac{E_b}{N_0}$, 256 QAM has a higher BER (bit error rate) than its counterparts, such as QPSK and BPS[2]. Nonetheless, its use in this project is justified because the 250MHz channel bandwidth it creates lends itself well to a wide range of commercially available bandpass filters. This practicality is in stark contrast to using a QPSK system, which would result in a 1000 MHz Bandwidth, making the procurement of practical BPFs very difficult.
256 QAM is the modulation technique adopted to create the 802.11ac, the next-generation wireless standard.

### 4. Results

The models were built and simulated using Agilent ADS, and the components used are available for purchase from the market. Each component's specification is outlined in the Appendix.

### 4.1 Power Analysis

The following figure shows a detailed model of the SE5003L1-R PA used. From the datasheet, the P1dB of the amplifier corresponds to a Gain of 32dB; from the graphs below, at the 1-dB compression point, the gain is approximately 31dB, and the output power is approx. 32dB, which corresponds to the published result. Secondly, the difference between the 1st and 3rd harmonics is approximately 11 dB, which corresponds to the literature formulae [1]. The discrepancy between the simulated and the practical model is less than 2%, which verifies the model's accuracy.

FROM PA MODEL:
$OIP3 = P1dB + 10.6 = 32dB + 10.6 = 42.6\ dB$

FROM SIMULATION:
$OIP3 = P1dB + 10.6 = 30.946 + 10.6 = 41.546\ dB$

The OIP3 of the actual PA model and the simulated model is approximately equal ($< 2.5\%$ difference) and, therefore, further justify the PA's accuracy.

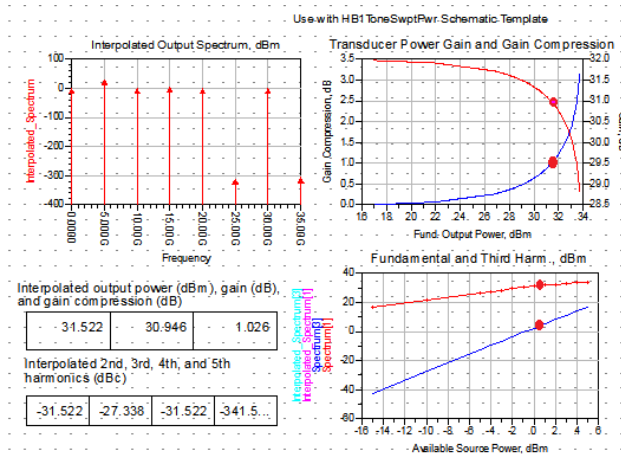

**Figure 4.** Simulated Power Amplifier.

From the Appendix and datasheet of the PA, it was stated that the output power of the PA is consistent over time with small variances. From the Power spectrum chart below, this statement is verified.

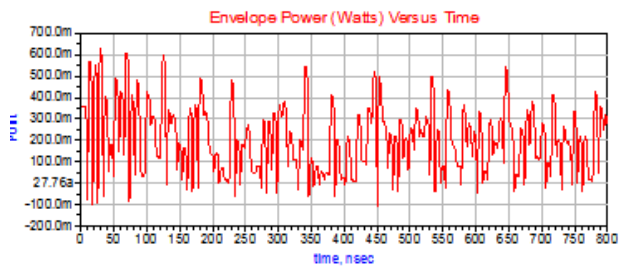

**Figure 5.** Power Spectrum of Transmitter.

Figure 6 shows the Transmitter input and output power, spectrum, and EVM at the antenna connector. From the spectrum, it can be determined that the null-to-null bandwidth is 250MHz, and the null frequency is 125 MHz per FCC regulations; unlicensed operation in the 5GHz band must adhere to a maximum transmission power of 250mW(24dBm)[2]. The transmission power obtained by this design is 23.31 dBm, which is below the maximum level and thus conforms to FCC regulations.

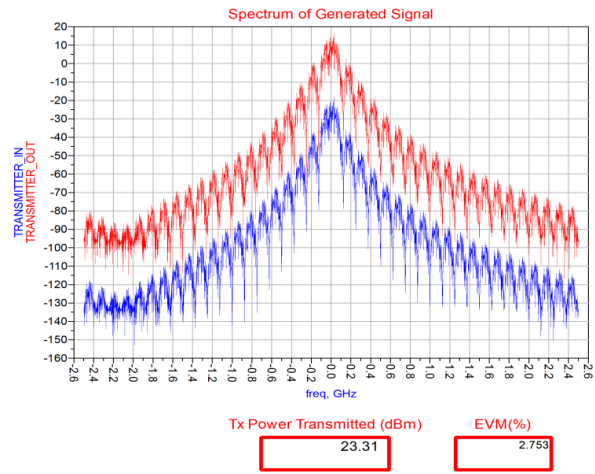

**Figure 6.** Modulation Spectra.

The EVM for this modulation was approximately 2.75 %. This deviation from the input constellation can be seen in the output constellation. The required EVM of 2% was not met in part because of the noise and spurious signals emitted from the power amplifier and the filter used. This could suggest that the devices used were somewhat defective. Also, the higher bit rate generated by 256 QAM would naturally generate a higher bit error rate. Given that this constellation is obtained without error correction schemes, it can be surmised that it is an acceptable modulation scheme.

### 4.2 Sensitivity and Bit Error Rate Analysis

The total Gain and Noise Figure (NF) for the Receiver was determined using SysCalc, where the total noise factor is,

$$F = F_1 + \frac{F_2-1}{G_1} + \frac{F_3-1}{G_1*G_2}$$
$$NF(dBm) = 10\log(F)$$

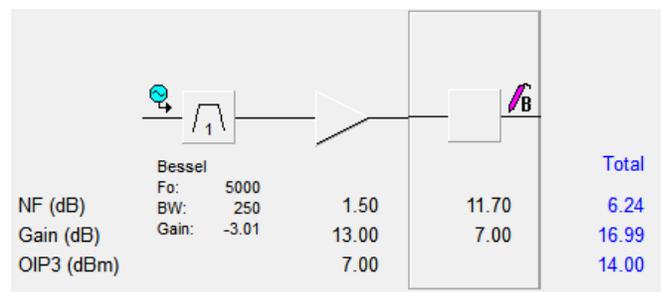

**Figure 7.** SysCalc Receiver Schematic.

From the waterfall chart for 256 QAM [4], a BER of $1 \times 10^{-5}$ results in:

$$\frac{E_b}{N_o} = 23.39 dB$$
$$SNR(dB) = \frac{E_b}{N_o}(dB) + 10logR - 10logB$$
$$SNR(dB) = 23.39 + 10\log(1e9) -$$
$$10\log(250e6) = 29.41 dB$$

@270K (*assuming room temperature conditions*)

$$Sensitivity(dBm) = -174dBm/Hz + F(dB) +$$
$$10\log(B) + SNR(dB)$$
$$= -174 db/Hz + 6.24 dB +$$
$$10\log(250e6) + 29.41 dB$$
$$= -54.37 dBm$$

Therefore, the $Receiver\ Sensitivity(dBm) = -54.37 dBm$

A Receiver's sensitivity is the minimum signal/power level it can detect. Since the power at our Receiver was -28.2dBm, higher than the sensitivity, the received power is within the acceptable range for this design.

### 4.3 Maximum Transmission Distance

From the FRIIS transmission equation:
$$P_{rx} = P_{tx} \times Grx \times Gtx \times \frac{\lambda^2}{4\pi R^2}$$
where,
$P_{rx} = -28.2 dBm = 1.51 \times 10^{-6} W$
$P_{tx} = 23.31 dBm = 0.214289 W$
$G_{rx} = G_{tx} = 0 dB\ (isotropic\ antennas) = 1$
$\lambda = \frac{c}{f} = \frac{3e8}{5e9} = 0.06 m$
$R_{max} = \frac{1}{4\pi}\sqrt{\frac{P_{tx}\lambda^2}{P_{rx}}} = \frac{1}{4\pi}\sqrt{\frac{0.214289*0.0036}{1.51\times10^{-6}}} = 1.79\ meters$

Therefore, the **maximum transmission distance was calculated to be 1.79 meters.**

Figures 8 and 9 below show the Modulation spectra before and after the amplifier at Transmitter and Receiver.

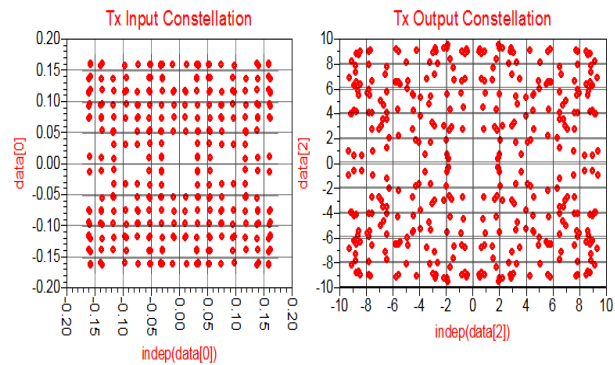

**Figure 8.** Transmitter Constellation.

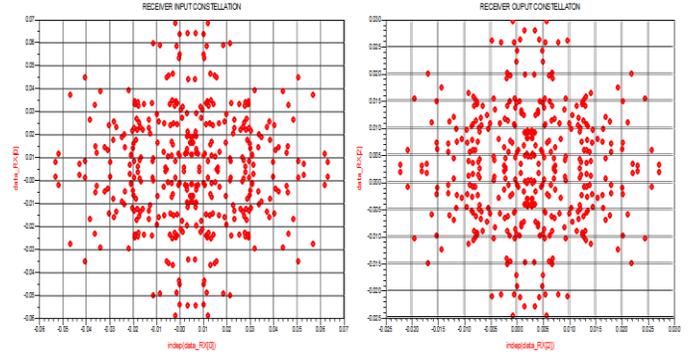

**Figure 9.** Receiver Constellation.

Figure 6 shows the input constellation expected for 256 QAM. However, due to the 90-degree phase change in the modulator and noise in the PA, some symbols were stacked on each other and distorted for the input and output constellation.

It was also observed that when the transmit power approached the saturation point of the PA, the symbols in the constellation diagram entered a circular shift. Therefore, the power transmitted was kept several orders below the P1dB (32dB) of the PA.

The receiver constellation, however, has a very similar input and output pattern, which is what was expected since a low noise amplifier is being used; thus, the symbols would not be as distorted. Both constellations at the Receiver also closely match the pattern at the Transmitter output. This means that most of the symbols sent by the transmitter were received by the Receiver with slight attenuation and symbol error. Despite the slight deviation between the Receiver's input and output constellations, the pattern is sufficiently informative to be demodulated.

The discrepancies in the maximum distance calculated could be attributed to the fact that the transmitter and Receiver had no gains at the antenna, and thus the signals would not have sufficient directivity and amplification on transmission and reception.

### 5. Conclusion

From the above simulations and analysis, it can be observed that the Homodyne architecture is a very simple and cost-effective transceiver system, which makes it very appropriate for M-ary modulation schemes such as 256 QAM. However, due to its simplicity and the higher bit error rates associated with 256 QAM, additional error correction schemes, more tightly controlled filters, and adaptive amplifiers may be necessary to obtain EVMs less than the accepted 2% level.

## A. Bill of Materials

### A.1 Transmitter

1) Power Amplifier (SE5003L1-R)
Freq (GHz):     5.15 – 5.90 GHz
Supply (V):     3 - 5
Current (mA):   205 @ Pout = +19 dBm
Gain (dB)       32
Output P1dB (dBm)    32
Input Return Loss    -11
$2^{nd}$ $3^{rd}$ harmonics    -45
Pout (@5.0V ) =    +19dBm, EVM <1.8%
Cost:    $4.40

2) ADL5375 (Quadrature Modulator)
Output frequency range: 400 MHz - 6 GHz
P1 dB output:    >9.4 dBm (450 MHz - 4 GHz)
Output Return loss:    < 12 dB (450 MHz - 4.5 GHz)
Noise floor:    −160 dBm/Hz @ 900 MHz
Sideband suppression:    <−50 dBc @ 900 MHz
Carrier feedthrough:    <−40 dBm @ 900 MHz
IQ3dB bandwidth:    ≥ 750 MHz
ADL5375-05:    500 mV
ADL5375-15:    1500 mV
Single supply:    4.75 V - 5.25 V
Cost:    $10.30

3) Bessel Filter (TGB2010-EPU-SM)
Frequency Band (GHz): 5 - 9
Gain (dB):    -3
Group Delay:    <±2.0ps to $F_o$
Return Loss:    15dB
Filter Bandwidth:    ± 0.125 GHz
Cost:    $2.79

### A.2 Receiver

1) LNA (SKY65981)
Freq (GHz):     5.15 – 5.85 GHz
Supply (V):     0 – 3.6
Current (mA):   205 @ Pout = +19 dBm
Gain (dB)       13
Noise Frequency (dB) 1.5
OIP3 (dB)    7
Output P1dB (dBm)    0
Input Return Loss    9
Cost: $3.78

2) Bessel Filter (TGB2010-EPU-SM)
Frequency Band (GHz): 5 - 9
Gain (dB): -3
Group Delay: <±2.0ps to $F_o$
Return Loss: 15dB
Filter Bandwidth: ± 0.125 GHz
Cost: $2.79

3) ADL5380 (Quadrature Demodulator)
   RF and LO frequency: 400 MHz - 6 GHz
   Input IP3:30 dBm @ 900 MHz
      28 dBm @1900 MHz
   Input IP2:   >65 dBm @ 900 MHz
   IP1dB:    11.6 dBm @ 900 MHz
   Voltage conversion gain: ~7 dB
   Demodulation bandwidth: ~390 MHz
   Noise figure:
      10.9 dB @ 900 MHz
      11.7 dB @ 1900 MHz
   Phase accuracy: ~0.2°
   Amplitude balance: ~0.07 dB
   Baseband I/Q drive: 2 V p-p into 200 Ω
   Single 5 V supply
Cost: $10.30